# Modeling Emotional Dynamics in Social Networks: Uncovering the Positive Role of Information Cocoons in Group Emotional Stabilization

Jinhu Ren[1], Xifei Fu[2], Tianlong Fan[1,*] and Linyuan Lü[1,*], *Senior Member, IEEE*

*Abstract*—Information cocooning—amplified by algorithmic filtering—poses complex challenges for emotional dynamics in online social networks. This study explores how algorithmically reinforced information cocooning shapes information diffusion and group emotional dynamics in online social networks. We propose a viewpoint-based network evolution model that simulates structural transformations driven by user preferences. To model the hidden influence of personalized comment recommendations, we introduce the Hidden Comment Area Cocoon (H-CAC)—a novel higher-order structure that captures cocooning at the comment level. This structure is integrated into an emotion spreading model, enabling the quantification of how cocooning affects collective sentiment. By defining Recommendation Accuracy (RA) as a tunable parameter, we systematically evaluate its impact on emotional volatility and polarization. Extensive simulations, validated with real-world data, reveal that while cocooning reduces content diversity, it can significantly enhance emotional resilience within groups. Our findings offer a new computational lens on the dual role of cocooning and provide actionable insights for designing emotionally stable, algorithmically governed social platforms.

## I. INTRODUCTION

Online social media has emerged as a primary platform for information dissemination [1, 2], allowing users to access and share vast amounts of content rapidly [3, 4]. Concurrently, social interactions on these platforms often involve complex emotional activities, including identification and empathy. However, information dissemination in such environments is neither entirely random nor unbiased. It is shaped significantly by individual preferences, selective interests, opinions [5, 6], algorithmic recommendations, and collective social influence [7, 8]. Thus, predicting information spreading trends under the intervention of recommendation technology, and exploring how these processes shape individual and group emotional dynamics has become a critical research issue [9-12]. Emotional contagion, a prevalent phenomenon in online social networks, describes how emotional states can be transmitted from one individual to another through social interactions, symbolic gestures, or textual communications [13, 14]. The dynamics of emotional contagion encompass multiple stages, including emotional elicitation, expression, and diffusion, which collectively depend on diverse factors such as individual psychology, social relationships, technological mediation, and environmental contexts [15, 16]. Previous studies emphasize emotional contagion as more than individual-level emotional exchanges, rather, it significantly influences collective social behaviors, community health, and even macro-level societal stability [17].

Various studies have provided empirical validation for emotional contagion and its mechanisms in social networks from various aspects. Kramer [18] highlighted emotional contagion's ability to propagate not only through direct textual communication, but also via indirect media. Kramer et al.'s large-scale experiment on Facebook [19] further confirmed that emotional shifts could occur implicitly, without user awareness. Furthermore, research by Stieglitz and Dang-Xuan [20] demonstrated that emotionally charged messages spread more rapidly and extensively compared to neutral ones. Hong et al. [21] proposed a Personalized Virtual and Physical cyberspace-based Emotional Contagion Model (PVP-ECM), utilizing mean field theory to realistically simulate individualized emotion spreading, demonstrating the significant role of personal traits in group emotional dynamics.

Recent studies have expanded emotional contagion's practical applications into diverse fields. For instance, Zhang et al. [22] proposed a dual-process emotional contagion mechanism in tourism scenarios and verified the model of emotion contagion, which provides a basis for the development of sustainable tourist-host relationship. Similarly, Shang et al. [23] proposed a game-theoretic evacuation model incorporating emotional contagion, examining its impacts on crowd evacuation efficiency. These studies collectively underscore the importance of elucidating emotion propagation mechanisms to effectively comprehend and manage collective behaviors in social systems.

Another critical phenomenon related to emotional dynamics in online social media is the "information cocoon". This term refers to the socially insulated environments wherein users selectively consume information aligned with personal preferences, interests, biases, or pre-existing beliefs, resulting in cognitive isolation and selective perception biases [24]. Modern-day information cocoons primarily result from personalized recommendation algorithms and group influence of social media platforms [25, 26].

[1] School of Cyber Science and Technology, University of Science and Technology of China, Hefei 230026, China.
[2] School of Computer and Communication, Lanzhou University of Technology, Lanzhou 730050, China.
* Corresponding authors: Tianlong Fan (tianlong.fan@ustc.edu.cn); Linyuan Lü (linyuan.lv@ustc.edu.cn).

Research on information cocoons has predominantly focused on their formation mechanisms and their negative consequences. Piao et al. [27] explored how human–AI adaptive interactions create information cocoons that exacerbate societal polarization. Zhang et al. [28] observed that reduced costs and user churn associated with algorithmic recommendations encourage users' passive acceptance, sustaining long-term cocoon environments. Hou et al. [29] quantitatively demonstrated that recommendation algorithms based on content similarity severely limit information diversity and system navigability, highlighting the harmful effects of information cocoons.

Recognizing these adverse effects, some researchers have proposed approaches for mitigating information cocoons. Lin et al. [30] proposed a Feature Disentanglement Self-Balancing (FDSB) re-ranking method to diversify recommendation outputs, thus preventing cocoon effects. Ni et al. [31] introduced the Equilibrium of Individual Concern-critical Influence Maximization (EIC-cIM) model, which aims to balance personal interests and curtail the negative impacts of information cocooning during dissemination processes.

Nevertheless, existing research predominantly views information cocoons as inherently negative, largely overlooking potential positive effects under specific conditions. Information cocoons may satisfy psychological needs by offering comfort, identity reinforcement, and emotional support [32], thus potentially promoting emotional stability and social cohesion. Therefore, recognizing the inevitability and persistence of information cocoons, it is valuable to explore whether and how this phenomenon can positively affect emotional dynamics at the collective level.

Motivated by these insights, this study quantitatively explores the interplay between information cocooning and emotional contagion in online social networks. Specifically, we first propose a network evolution model incorporating individual viewpoints, aiming to realistically simulate structural evolution during information diffusion. Next, we introduce the Recommendation Accuracy (RA) index, enabling quantitative assessment of recommendation algorithms' effectiveness. Based on RA, we further develop a Hidden Comment Area Cocoon (H-CAC) structure derived from higher-order network theory, accurately capturing cocooning effects generated through comment interactions. Leveraging this innovative H-CAC framework, we build an emotion spreading dynamics model designed to quantify how information cocoon structures influence group emotional fluctuations. Through comprehensive experimental validation with real-world datasets and robustness analyses, we demonstrate the nuanced positive roles that information cocooning can play in stabilizing group emotions.

This study contributes to the existing literature in several significant ways.

- Unlike traditional studies focusing exclusively on the negative dimensions of information cocooning, our findings reveal conditions under which cocooning positively contributes to group emotional stability.
- We introduce and rigorously define the novel H-CAC structure, explicitly capturing the hidden recommendation dynamics present in online comments.
- By proposing the RA index, we enable precise control and analysis of algorithmic recommendations within our simulation models.

Recognizing the practical challenges of entirely eliminating information cocoons, our work advocates a nuanced approach—mitigating their harmful effects while strategically leveraging their potential benefits for emotional stability and societal cohesion.

The rest of this paper is organized as follows: Section II describes the modeling framework in detail. Specifically, in Part A, we propose a network evolution model incorporating individual viewpoints to capture realistic network dynamics. In Part B, we introduce the H-CAC framework, which explicitly models the cocoon structures formed through comment recommendations. In Part C, we further establish an emotion spreading dynamics model based on the H-CAC framework to analyze and quantify group-level emotional fluctuations. Section III presents comprehensive simulations results and analysis. Part A discusses simulation outcomes, highlighting the emotional spreading dynamics under various scenarios. Part B validates the proposed model through comparisons with real-world data, evaluating its accuracy and identifying strengths and limitations. Part C performs sensitivity analysis to quantitatively evaluate the influence of RA on group emotion stability. Part D evaluates the robustness and generalization capability of the proposed model across diverse network conditions. Finally, Section IV concludes the paper by summarizing key findings and discussing the theoretical contributions and practical implications of this research.

## II. MODEL BUILDING

This study introduces several basic assumptions to simplify the modeling framework and define clear logical boundaries. These assumptions will guide the experimental design, data collection, and result analysis.

- Assumption 1: $RA \in [0,1)$ denotes the global accuracy level of recommendation algorithms. In realistic sense, the RA represents the accuracy of the smart algorithms of a social platform to make a push to the user. A higher $RA$ value indicates greater recommendation accuracy.
- Assumption 2: Repeated comments diminishing effects on user sentiment; thus, repeated comments between the same pair of nodes are not considered.

*A. Network Evolution Modeling Based on Individual Viewpoints*

We begin by constructing an empty network $G(V, \emptyset)$, where $V = \{v_1, v_2, \cdots, v_N\}$ denotes the set of nodes. Each node $v_i$ possesses three intrinsic attributes: a viewpoint attribute $p \in (0,1)$, an emotion attribute $m \in (-1,1)$, and a faith attribute $f$. The probability density functions for these attributes are defined as follows:

# Modeling Emotional Dynamics in Social Networks:
## Uncovering the Positive Role of Information Cocoons in Group Emotional Stabilization

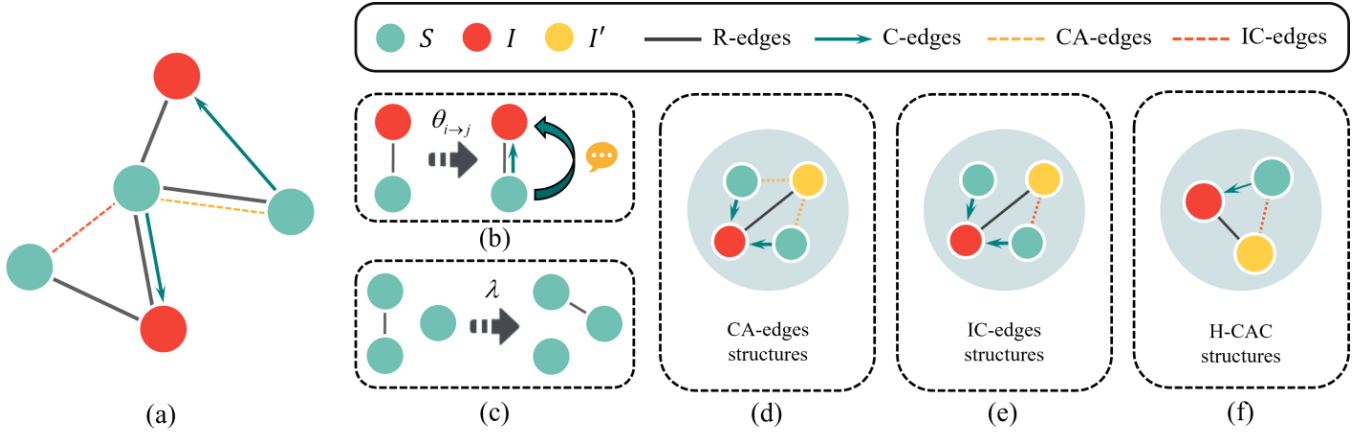

**Fig. 1** | Schematic illustration of the network model and core structural concepts. (a) The hybrid network model consists of unpublished nodes $S$ (green), published nodes $I$ (red), and temporarily transitioning nodes $I'$ (yellow). It includes two types of real edges: relationship edges (R-edges, black solid lines) and comment edges (C-deges, green arrows), as well as two types of virtual edges: comment area edges (CA-edges, yellow dashed lines) and information cocoons edges (IC-edges, red dashed lines). (b) The generation of directed comment edges based on the probabilistic mechanism $\theta_{i \to j}$. (c) Rewiring of relationship edges with probability $\lambda$ guided by viewpoint similarity. (d) Construction of CA-edge structures, forming fully connected virtual links from commenters to candidate nodes. (e) Filtering of CA-edges by recommendation accuracy (RA) to generate IC-edges, capturing algorithmically induced filtering effects. (f) Formation of higher-order Hidden comment area cocoon (H-CAC) structures through the integration of IC-edges, reflecting implicit cocooning in comment spaces.

$$\begin{cases} F(p), F(f) = \frac{4}{\sqrt{2\pi}} exp(-8(x-\frac{1}{2})^2) \\ F(m) = \frac{1}{\sqrt{2\pi}} exp(-\frac{x^2}{2}) \end{cases}. \quad (1)$$

Subsequently, directed relationship edges representing interpersonal relationships are added to the network $G$ according to the Barabási–Albert (BA) scale-free network model [33]. These edges capture the paths through which information spreads between individuals. The adjacency matrix describing relationship edges is denoted as $E_r = [e_{ij}]_{N \times N}$.

During the spreading process, directed comment edges representing commenting behavior dynamically emerge. We denote the adjacency matrix for these comment edges by $E_c = [c_{ij}]_{N \times N}$, where $c_{ij}$ indicates that node $v_i$ has made a comment directed towards node $v_j$.

Each node can occupy one of two distinct states: the unpublished state $S$, indicating that the node has not yet spread information, and the published state $I$, indicating that the node has already disseminated information to its neighbors.

Based on the aforementioned framework, we establish a hybrid static network (as illustrated in Fig. 1(a)) comprising three node attributes $(p, m, f)$, two node states $(S, I)$ and two edge types (relationship edges and comment edges). Building upon this structure, we define dynamic evolution rules governing how the network evolves over time. This dynamic evolution includes two integral processes: the evolution of comment edges and the evolution of relationship edges.

The evolution of comment edges is illustrated schematically in Fig. 1(b). When a node $v_i$ disseminates information, a directed comment edge from node $v_i$ to node $v_j$ emerges with probability $\theta_{i \to j}$, given by:

$$\theta_{i \to j} = \theta_0 \cdot \frac{f_j + (1-2|p_i - p_j|)^2}{2}, \quad (2)$$

where $\theta_0$ is a scaling parameter, $f_j$ accounts for the influence of parent node $j$'s faith attribute, while the later term captures the commenting behavior arising from opinion polarization.

The evolution process for relationship edges is illustrated schematically in Fig. 1(c). During information dissemination, existing relationship edges are probabilistically rewired with probability $\lambda$. Specifically, the reconnection process utilizes the viewpoint similarity-based filtering procedure outlined in Algorithm 1. New edges are formed by randomly selecting nodes from the filtered recommendation list, which depends explicitly on the Recommendation Accuracy (RA) index defined previously.

This comprehensive network evolution modeling approach effectively integrates individual viewpoints, emotional attributes, and commenting behaviors, providing a realistic foundation for the subsequent analysis of information and emotion dynamics within online social networks.

### B. Hidden Comment Area Cocoon (H-CAC) Framework

We propose the Hidden Comment Area Cocoon (H-CAC) framework to explicitly model cocoon structures arising from algorithm-driven personalized recommendations in online comment sections. The underlying logic of this framework is derived from two key observations. Firstly, the comment area beneath any given post comprises comments reflecting the commenters' individual viewpoints. Secondly, modern social platforms deploy intelligent recommendation algorithms that present personalized comment sections, causing users with different preferences to see different subsets of comments. Conse

Modeling Emotional Dynamics in Social Networks:
Uncovering the Positive Role of Information Cocoons in Group Emotional Stabilization

**Algorithm 1:** Viewpoint Similarity Algorithm Based on Recommendation Accuracy.

```
Input:
 1: G ← (V, E)         // Social network graph structure
 2: v_pre ∈ V          // Pre-perturbed Nodes
 3: RA ∈ [0,1]         // Recommendation accuracy parameter
Output:
 4: v_post ⊆ V         // Post-perturbed Nodes
Procedure:
 5: // Phase 1: Viewpoint comparison
 6: differences ← []
 7: for each v_i in V do
 8:     if v_i ≠ v_pre then
 9:         v_t ← G.nodes[v_pre].viewpoint
10:         v_c ← G.nodes[v_i].viewpoint
11:         Δ ← |v_t - v_c|
12:         differences.append((Δ, v_i))
13:     end if
14: end for
15: // Phase 2: Ranking selection
16: sort_ascending(differences)
17: total ← len(differences)
18: k ← total × (1 - RA)
19: v_post ← ∅
20: for i in 0 to k-1 do
21:     v_post.add(differences[i]. v_i)
22: end for
23: return v_post
```

quently, these personalized recommendations create hidden cocoon structures in the comment areas, shaping users' perceptions of majority opinions in subtle yet impactful ways.

To precisely characterize this phenomenon, we define two categories of virtual edges based on the hybrid network structure established previously (as illustrated in Fig. 1(a)).

Initially, comment area edges (CA-edges) are constructed to represent the implicit interactions within the comment section realistically. Specifically, for each node $I_i$, we define its set of potential targets for spreading information as $I_i'$. Consider a node $I_{ij}' \in I_i'$, which represents a potential target node for information dissemination. We introduce virtual edges connecting $I_{ij}'$ with all nodes commenting on the node $I_i$. These edges are termed comment area edges (CA-edges), depicted in Fig. 1(d), and the commenters are collectively denoted by the node set $N_{CA}$. To ensure a fair representation free from selection biases, the set of CA-edges, denoted as $E_{CA}$, is determined based on viewpoint similarity rankings between commenters and the target node $I_{ij}'$.

Building upon the CA-edge structure, we further introduce Information Cocoon edges (IC-edges), which explicitly model the filtering effect introduced by the Recommendation Accuracy (RA), $\delta \in [0,1)$, as shown in Fig. 1(e). RA effectively constrains the visibility of comments by limiting the range of nodes whose viewpoints are sufficiently similar to the target node. Formally, we define the set of IC-edges, $E_{IC}$, through the filtering relation:

$$E_{IC} = E_{CA} \cdot (1 - \delta). \tag{3}$$

Consequently, an implicit higher-order information cocoon emerges, illustrated schematically in Fig. 1(f). We formally denote this higher-order structure as $I_{ij}' \xrightarrow{N_{IC}(i,j)} I_i$. Here, $N_{IC}$ denotes the set of filtered comment nodes forming the resulting cocoon structure.

*C. H-CAC-based Modeling of Emotion Spreading Dynamics*

To quantitatively explore group-level emotional dynamics influenced by information cocoon structures, we develop an emotion spreading dynamics model grounded in the established H-CAC framework. Our modeling approach integrates principles from classical infectious disease dynamics, specifically employing the susceptible-infected (SI) paradigm [34]. In the SI model, nodes permanently remain in the "infected" state upon initial exposure, analogous to how persistent information dissemination affects individuals on online platforms. Correspondingly, in our proposed model, information released initially by a subset of individuals ("infected" nodes) propagates through the network, continuously impacting group emotions.

Within our network, nodes with information are defined as "known" nodes $I$, while nodes yet to receive information are termed "unknown" nodes $S$. At time $t$, we denote their quantities as $S(t)$ and $I(t)$, respectively, satisfying $S(t) + I(t) = N$. The normalized densities of unknown and known nodes at time $t$ are given by:

$$\begin{cases} s(t) = \frac{S(t)}{N} \\ i(t) = \frac{I(t)}{N} \end{cases}, \text{ with } s(t) + i(t) \equiv 1. \tag{4}$$

The fundamental dynamics of information spread can then be represented as:

$$\begin{cases} \frac{ds}{dt} = -\alpha s i \\ \frac{di}{dt} = \alpha s i \end{cases}, \tag{5}$$

where $\alpha$ denotes the spreading rate.

Considering the realistic "Matthew effect" in social networks—where nodes with higher degrees exert disproportionately greater influence—we extend our spreading rate definition to reflect nodes' degrees and their faith attribute $f$. For node $v_i$ having degree $k_i$, the spreading rate is formulated as:

$$\alpha_i = \frac{\alpha_0}{2}\left(f_i + \frac{k_i - \langle k \rangle}{k_{max}}\right), \tag{6}$$

where $\alpha_0$ is the base spreading rate, $\langle k \rangle$ is the global average node degree, and $k_{max}$ is the maximum degree in the network. Thus, the dynamics of information spread follow a logistic form:

$$\frac{di}{dt} = \alpha_0 \cdot \frac{1}{2}(f_i + \frac{k_i - \langle k \rangle}{k_{max}}) \cdot i(1 - i). \tag{7}$$

Solving analytically yields the logistic growth solution:

$$i(t) = \frac{i_0 e^{\frac{\alpha_0}{2}\left(f_i + \frac{k_i - \langle k \rangle}{k_{max}}\right)t}}{1 - i_0 + i_0 e^{\frac{\alpha_0}{2}\left(f_i + \frac{k_i - \langle k \rangle}{k_{max}}\right)t}}, \text{ where } i_0 = i(0). \tag{8}$$

Leveraging the established H-CAC framework $I_{ij}' \xrightarrow{N_{IC}(i,j)} I_i$, we formulate an emotional perturbation function describing emotional adjustments at each spreading event. For a given node $I_{ij}'$ subject to spreading and having $l$ connected H-CAC structures, the emotional perturbation $\Delta m_{ij}$ is defined as:

$$\Delta m_{ij} = \frac{1}{l}\sum_{k=1}^{l}\left(\frac{m_i + m_k}{2} - m_{ij}\right), \tag{9}$$

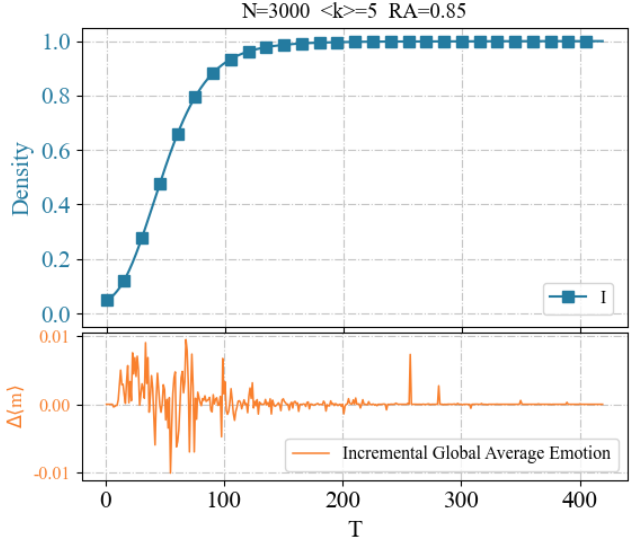

**Fig. 2.** | Temporal evolution of spreader density and group emotional fluctuation. The upper panel shows the rise in spreader density I over time T under a network configuration of $N$=3000, $\langle k \rangle$=5, and RA=85%. The lower panel illustrates the corresponding incremental change in global average emotion $\Delta \langle m \rangle$, highlighting strong fluctuations during the early stage of information diffusion.

---

**Algorithm 2:** Emotion spreading algorithm based on comment area cocooning

**Input:**
1: $G \leftarrow (V, E)$  // Social network graph
2: $i\_0 \in (0,1)$  // Initial spreader density
3: $alpha \in [0,1]^{|V|}$ // Node activation probability vector
4: $Theta \in [0,1]$  // Comment edge threshold
5: $Lambda \in [0,1]$  // Edge reconnection probability

**Output:**
6: $G \leftarrow (V, E')$  // Updated emotion propagation network

**Procedure:**
7: spreaders ← random_select(V, floor(i_0 * |V|))
8: // Phase 1: Information diffusion
9: for each I in spreaders do
10: |  for each neighbor u of I do
11: |  |  if rand() < alpha[I] then
12: |  |  |  activate(u)
13: |  |  |  if rand() < Theta then
14: |  |  |  |  E.add_edge(I, u, type='comment')
15: |  |  |  |  l ← I.in_degree()
16: |  |  |  |  sum_term ← Σ_{k=1}^l [(m_I + m_k)/2 - m_IJ]
17: |  |  |  |  m_IJ += (1/l) * sum_term   // Eq.(10)
18: // Phase 2: Topology update
19: for each edge (x,y) in E do
20: |  if rand() < Lambda then
21: |  |  y_new ← Algorithm_1(G, x)
22: |  |  E.remove_edge(x, y)
23: |  |  E.add_edge(x, y_new)
24: return G

where $m_i$, $m_k$ and $m_{ij}$ represent emotional states of nodes $I_i$, $N_{IC}$, and $I'_{ij}$, respectively. Thus, the emotional dynamics equation governing the node during information spreading is updated as:

$$m_{ij} \leftarrow m_{ij} + \frac{1}{l}\sum_{k=1}^{l}\left(\frac{m_i+m_k}{2} - m_{ij}\right). \quad (10)$$

For analysis purposes, the global average emotion can be simply stated as:

$$\langle m \rangle = \frac{1}{N}\sum_{i=1}^{N} m_i. \quad (11)$$

Thus, between each time step, the change in global average mood can be expressed as:

$$\Delta\langle m \rangle_t = \langle m \rangle_t - \langle m \rangle_{t-1}, t \geq 1. \quad (12)$$

This systematic formulation (summarized in Algorithm 2) rigorously links network topology, information cocoon structure, and emotional dynamics, providing a robust analytical foundation for understanding the nuanced interplay between personalized recommendation mechanisms and collective emotional responses in online social networks.

### III. SIMULATION AND ANALYSIS

This section provides an in-depth analysis of simulation outcomes based on the proposed model and validates these outcomes against real-world data, aiming to ensure the robustness and applicability of our theoretical framework. The experimental network employed in this analysis comprises $N = 3000$ nodes, with an average node degree $\langle k \rangle = 5$, and a $RA$ set at 85%. A subset of nodes is initially designated as spreaders, representing the sources from which information dissemination begins. Key parameter settings are detailed in TABLE 1.

**TABLE 1**
**Parameter Setup**

| Symbol | Value | Definition |
|---|---|---|
| $N$ | 3000 | The total number of nodes |
| $k$ | 5 | Average degree |
| $i_0$ | 0.006 | The initial retweeting density |
| $\lambda$ | 0.1 | Probability of reconnection of relationship edges |
| $\theta_0$ | $8.62 \times 10^{-3}$ | Scale control parameters of Eq.(2) |

*A. Simulation Results*

The simulations are executed using Monte Carlo methods. Fig. 2 illustrates the changes in spreader density and group emotions over time. Specifically, the blue curve tracks the evolving density of informed nodes (spreaders), while the orange curve indicates instantaneous change in global average emotions ($\Delta\langle m \rangle$). From $T = 0$ to 100, the density of spreaders quickly rises from near zero to approximately 0.9, accompanied by significant emotional fluctuations. This simultaneous pattern suggests a direct correlation between rapid information dissemination and pronounced group emotional volatility. Such scenarios often arise during the rapid spread of controversial or emotionally charged information, potentially leading to cognitive overload, confusion, or emotional distress among individuals.

Complementing the information diffusion dynamics, our model simultaneously generates a directed comment network (Fig. 3(a)), representing the interactions occurring via comments. Nodes with higher incidences (large in-degree) appear prominently, indicating heightened comment activity directed towards these individuals. Conversely, numerous isolated nodes reflect "silent nodes" that neither comment nor receive comments during the dissemination process. Further visualization in Fig. 3 (b) shows emotion distribution across nodes at $RA = 85\%$, highlighting neutral emotions dominating among influential nodes. Reducing $RA$ to 20% (Fig. 3 (c)) increases emotional diversity significantly, implying that higher recommendation accuracy fosters neutral emotional conformity among influential individuals, albeit reducing informational diversity.

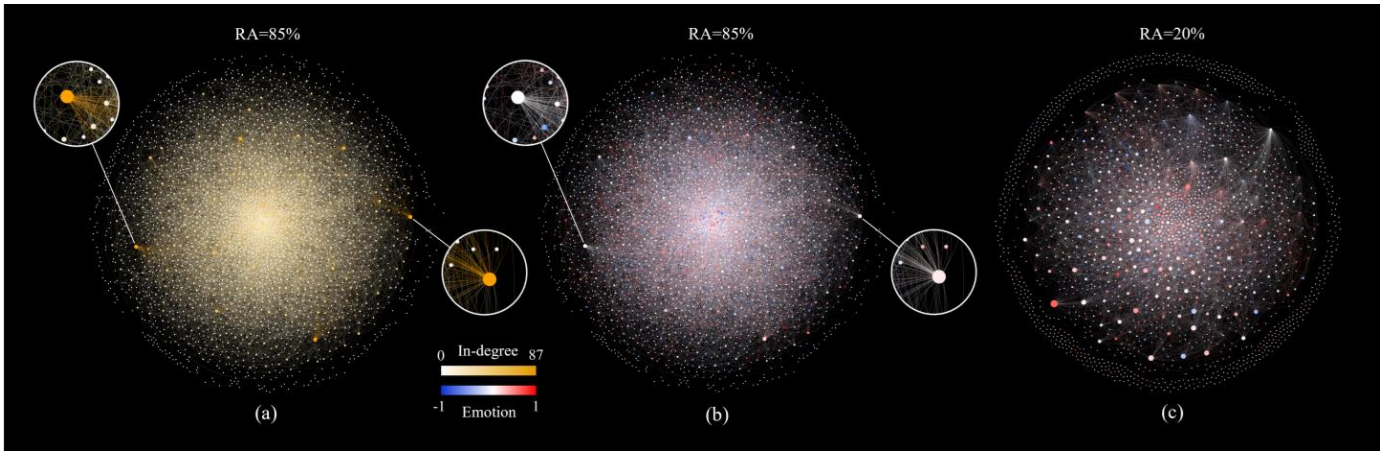

**Fig. 3.** | Structural and emotional features of the comment network during diffusion under different RA levels. (a) Visualization of node in-degree distribution at $RA = 85\%$; nodes with higher in-degree are more frequently commented upon. (b) Corresponding emotion m's distribution of nodes at $RA = 85\%$. (c) Emotion distribution at $RA = 20\%$, revealing increased emotional polarization and reduced neutrality among high-degree nodes.

### B. Validation with Real-world Data

Given that many complex systems lack comprehensive theoretical guidance, empirical validation serves as a crucial tool for verifying model accuracy. Our simulations are compared with real-world datasets from Sina Weibo and Math Overflow [35], focusing on two critical dimensions: the trends of information spreading and the structure of generated comment networks.

We utilized three real-world Sina Weibo datasets:

- Case 1: Data from the event "A celebrity in a vehicle hit a camel", spanning approximately 14 hours.
- Case 2: Information spread related to "Renowned doctors of a hospital in Shanghai", covering roughly a month.
- Case 3: Data on "Copyright disputes on a platform", over a similar duration.

We normalized the data numerically and aligned them temporally to ensure that the observations were intuitive and concise. In Fig. 4, experimental simulations (blue curve) closely match real-world cases (orange, red, and green curves), particularly in the rapid initial spreading phase and eventual stabilization of information density. The close alignment validates the model's efficacy in realistically replicating spreading dynamics.

We further quantitatively examine the generated comment networks, comparing them to the Math Overflow temporal network dataset. To enhance analytical clarity, silent nodes from our experimental data were removed (Fig. 6(c)). After this adjustment, the experimental network closely approximates the real network regarding average node degrees and degree distributions.

Minor differences noted in the mid-degree region ($k \in (10,30)$) stem from the plausible developmental laws of networks from infancy to maturity. The social commenting networks we modeled are much earlier and more dynamic, and differ normally from the phenomenon of mature networks showing significantly heavy-tailed distributions [36]. This comparative analysis further confirms the model's reliability in simulating real-world comment dynamics.

### C. Sensitivity Analysis of RA on Group Emotion

To assess how recommendation accuracy (RA) influences emotional dynamics, we analyze the temporal evolution of the global average emotion $\langle m \rangle$ under different RA values. The results are shown in Fig. 5. Consistent with the findings in Fig. 2, significant emotional fluctuations are primarily observed during the early phase of diffusion. To reduce visual noise and highlight core trends, we further extract the statistical range of across time, as shown in TABLE 2.

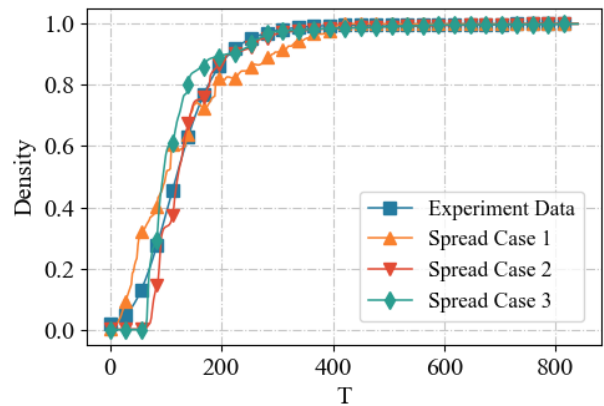

**Fig. 4** | Comparison between simulated and empirical spreading dynamics. The blue curve represents the simulation results from the proposed model. Orange, red, and green curves correspond to real-world cases (Case 1, Case 2, Case 3). The high degree of alignment confirms the model's capacity to capture realistic spreading patterns across diverse scenarios.

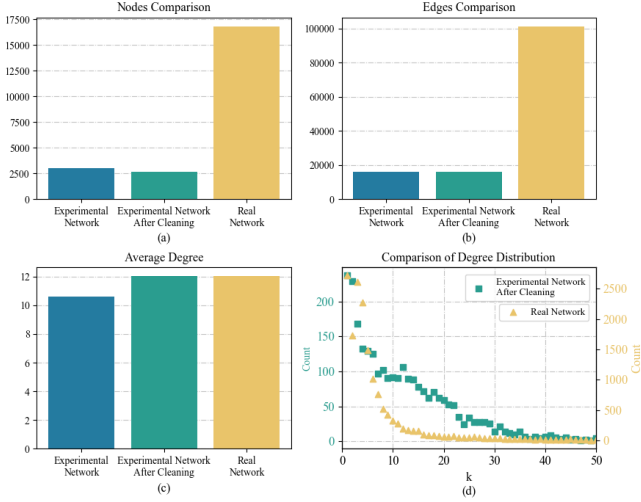

**Fig. 6** | Comparative analysis between the simulated and real-world comment networks. (a) Number of nodes. (b) Number of edges. (c) Average degree. (d) Degree distribution comparison.

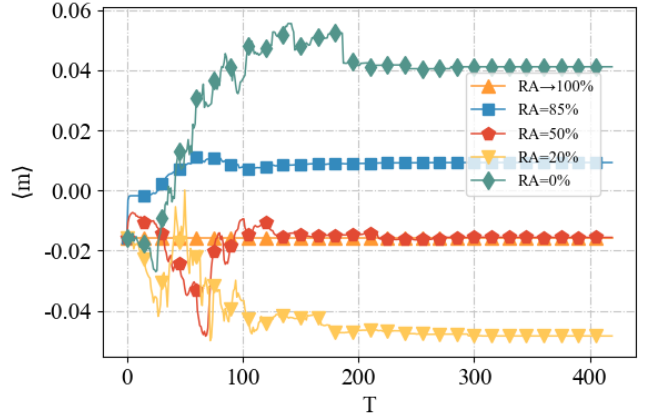

**Fig. 5** | Evolution of global average emotion under different levels of $RA$. Curves represent emotion dynamics over time for $RA \in \{0\%, 20\%, 50\%, 85\%, 100\%\}$. Higher $RA$ values correspond to lower emotional fluctuation, indicating that more accurate recommendations contribute to greater emotional stability at the group level.

All simulations are conducted under controlled initial network configurations to ensure comparability. Therefore, the initial values of $\langle m \rangle$ are identical across different RA settings. As indicated in TABLE 2, increasing RA systematically reduces the amplitude of emotional fluctuation. This suggests that higher RA contributes to greater emotional stability within the group.

This finding implies that when recommendation systems can more precisely align with users' cognitive preferences, the emotional volatility of users during information exposure tends to decrease. In other words, increased RA leads to less emotional polarization and a more harmonious public discourse environment. This insight offers a valuable angle for evaluating the psychological impacts of algorithmic design in social platforms.

In summary, our model successfully captures the nuances of information diffusion and emotion dynamics across diverse network environments, validated by rigorous comparison with empirical data. It confirms both the conceptual validity of the H-CAC structure and the practical accuracy of our generated comment networks, offering robust insights into real-world network dynamics.

*D. Model Adaptability Across Network Conditions*

To evaluate the generalizability of the proposed model, we conduct a series of adaptability experiments under varying network configurations. The results are presented in Fig. 7.

Fig. 7(a) examines the effect of network size. Despite scaling from small to large populations, the density of informed nodes over time remains stable, indicating the model's robustness in both micro- and macro-scale scenarios. Fig. 7(b) compares information spreading dynamics under two network types: the Watts-Strogatz (WS) small-world network [37] and the BA scale-free network. As expected, spreading progresses more slowly in WS networks due to fewer influential hubs. Nonetheless, both networks exhibit similar convergence trends, confirming the model's adaptability to different topologies. Fig. 7(c) explores different average degrees. As node connectivity increases, the diffusion accelerates marginally, but the overall dynamics remain qualitatively consistent. This reflects the model's ability to accommodate varying interaction densities in realistic settings. Fig. 7(d) evaluates the impact of different initial infection densities. Regardless of whether information originates from sparse or dense initial sources, the model consistently predicts realistic spreading trajectories. This indicates its effectiveness even in scenarios with incomplete early-stage monitoring.

Overall, these experiments demonstrate the proposed model's strong adaptability and generalization capacity, making it a viable tool for forecasting and managing information dissemination in diverse online environments.

TABLE 2
Global average emotion $\langle m \rangle$ under different RA values

| RA | Initial $\langle m \rangle$ | Minimum $\langle m \rangle$ | Maximum $\langle m \rangle$ | Difference of $\langle m \rangle$ |
|---|---|---|---|---|
| →100% | -0.01588 | -0.01588 | -0.01588 | 0.00000 |
| 85% | -0.01588 | -0.01588 | 0.01104 | 0.02692 |
| 50% | -0.01588 | -0.04846 | -0.00735 | 0.04111 |
| 20% | -0.01588 | -0.04997 | 0.00013 | 0.05009 |
| 0% | -0.01588 | -0.02704 | 0.05558 | 0.08261 |

\* Minimum and Maximum are the minimum and maximum values of $\langle m \rangle$ in the overall spreading process, respectively.

IV. CONCLUSION

This study presents a comprehensive model to explore how

Modeling Emotional Dynamics in Social Networks:
Uncovering the Positive Role of Information Cocoons in Group Emotional Stabilization

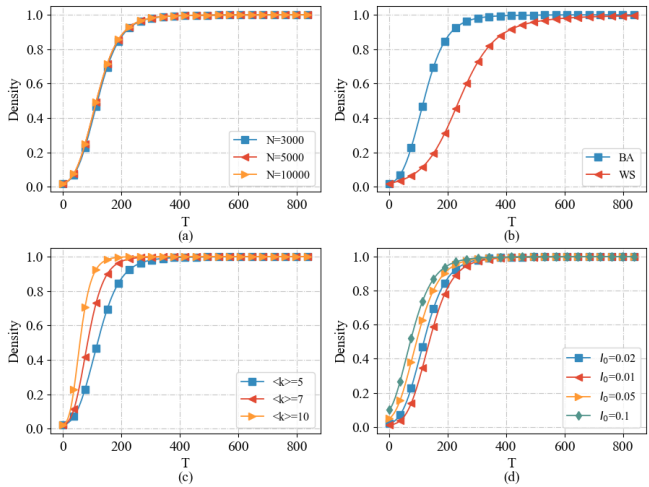

**Fig. 7** | Model adaptability across varying network conditions. Experiments on the generalization ability of the model are conducted in network environments of different (a) sizes, (b) network types, (c) average degree, and (d) initial infection densities.

algorithmically driven information cocoons influence emotional dynamics within online social networks. By integrating individual viewpoints into a network evolution model, we capture realistic user interactions under algorithmic mediation. The introduction of the RA parameter enables fine-grained control over recommendation strength, and the proposed H-CAC framework effectively represents the emergent cocooning structure formed through comment recommendations.

Our results demonstrate that while information cocooning is often perceived as a negative phenomenon, it may also stabilize group emotions under specific conditions. This duality suggests that, rather than aiming to eliminate cocoon effects entirely—a goal that is largely impractical in real-world environments—a more pragmatic approach is to manage their structure and influence. Our model supports this view by showing that higher RA levels, though limiting information diversity, can reduce group-level emotional volatility. The model's validity is reinforced through close alignment with real-world datasets, and its adaptability is confirmed under diverse network scenarios. This work not only advances theoretical understanding of information cocoons but also provides operational insights for designing emotionally resilient social media ecosystems.

## V. SIMULATION PLATFORM AND DATA SOURCES

Simulation experiments were implemented in Python using the PyCharm development environment (https://www.jetbrains.com/pycharm). Network visualizations were generated using Gephi (https://gephi.org/). Real-world information diffusion datasets were obtained from the WRD Big Data platform (https://research.wrd.cn/), used under license. Math Overflow temporal network data were sourced from the Stanford SNAP repository (https://snap.stanford.edu/data/sx-mathoverflow.html). Additional data access is available upon reasonable request and with proper authorization.

ACKNOWLEDGEMENTS

This work is supported by the National Natural Science Foundation of China (T2293771), the STI 2030—Major Projects (2022ZD0211400), the China Postdoctoral Science Foundation (Grant No. 2024M763131), the Postdoctoral Fellowship Program of CPSF (Grant No. GZC20241653), and the Sichuan Province Outstanding Young Scientists Foundation (2023NSFSC1919) and the New Cornerstone Science Foundation through the XPLORER PRIZE.

REFERENCES

[1] D. Brossard and D. A. Scheufele, "Science, new media, and the public," *science,* vol. 339, no. 6115, pp. 40-41, 2013.
[2] H. Gil de Zúñiga, N. Jung, and S. Valenzuela, "Social media use for news and individuals' social capital, civic engagement and political participation," *Journal of computer-mediated communication,* vol. 17, no. 3, pp. 319-336, 2012.
[3] A. M. Kaplan and M. Haenlein, "Users of the world, unite! The challenges and opportunities of Social Media," *Business horizons,* vol. 53, no. 1, pp. 59-68, 2010.
[4] F. Nian, J. Ren, and X. Yu, "Online Spreading of Topic Tags and Social Behavior," *IEEE Transactions on Computational Social Systems,* 2023.
[5] E. Bakshy, S. Messing, and L. A. Adamic, "Exposure to ideologically diverse news and opinion on Facebook," *Science,* vol. 348, no. 6239, pp. 1130-1132, 2015.
[6] D. A. Vega-Oliveros, L. Berton, F. Vazquez, and F. A. Rodrigues, "The impact of social curiosity on information spreading on networks," in *Proceedings of the 2017 IEEE/ACM International Conference on Advances in Social Networks Analysis and Mining 2017,* 2017, pp. 459-466.
[7] S. Flaxman, S. Goel, and J. M. Rao, "Filter bubbles, echo chambers, and online news consumption," *Public opinion quarterly,* vol. 80, no. S1, pp. 298-320, 2016.
[8] J. Ren, F. Nian, and X. Yang, "Two-Stage Information Spreading Evolution on the Control Role of Announcements," *IEEE Transactions on Computational Social Systems,* 2024.
[9] M. Fan, Y. Huang, S. A. Qalati, S. M. M. Shah, D. Ostic, and Z. Pu, "Effects of information overload, communication overload, and inequality on digital distrust: A cyber-violence behavior mechanism," *Frontiers in psychology,* vol. 12, p. 643981, 2021.
[10] W. Li, Y. Li, W. Liu, and C. Wang, "An influence maximization method based on crowd emotion under an emotion-based attribute social network," *Information Processing Management,* vol. 59, no. 2, p. 102818, 2022.
[11] K. Solovev and N. Pröllochs, "Moral emotions shape the virality of COVID-19 misinformation on social media," in *Proceedings of the ACM web conference 2022,* 2022, pp. 3706-3717.
[12] L. Geng, H. Zheng, G. Qiao, L. Geng, and K. Wang, "Online public opinion dissemination model and simulation under media intervention from different perspectives," *Chaos, Solitons&Fractals,* vol. 166, p. 112959, 2023.
[13] J. H. Fowler and N. A. Christakis, "Dynamic spread of happiness in a large social network: longitudinal analysis over 20 years in the Framingham Heart Study," *Bmj,* vol. 337, 2008.
[14] R. Pfitzner, A. Garas, and F. Schweitzer, "Emotional divergence influences information spreading in Twitter," in *Proceedings of the International AAAI Conference on Web and Social Media,* 2012, vol. 6, no. 1, pp. 543-546.
[15] M. De Choudhury, S. Counts, and M. Gamon, "Not all moods are created equal! exploring human emotional states in social media," in *Proceedings of the International AAAI Conference on Web and Social Media,* 2012, vol. 6, no. 1, pp. 66-73.
[16] L. Coviello *et al.*, "Detecting emotional contagion in massive social networks," *PloS one,* vol. 9, no. 3, p. e90315, 2014.
[17] R. M. Bond *et al.*, "A 61-million-person experiment in social influence and political mobilization," *Nature,* vol. 489, no. 7415, pp. 295-298, 2012.
[18] A. D. Kramer, "The spread of emotion via Facebook," in *Proceedings of the SIGCHI conference on human factors in computing systems,* 2012, pp. 767-770.
[19] A. D. Kramer, J. E. Guillory, and J. T. Hancock, "Experimental evidence of massive-scale emotional contagion through social networks,"

*Proceedings of the National academy of Sciences of the United States of America,* vol. 111, no. 24, p. 8788, 2014.

[20] S. Stieglitz and L. Dang-Xuan, "Emotions and information diffusion in social media—sentiment of microblogs and sharing behavior," *Journal of management information systems,* vol. 29, no. 4, pp. 217-248, 2013.

[21] X. Hong, G. Zhang, D. Lu, H. Liu, L. Zhu, and M. Xu, "Personalized crowd emotional contagion coupling the virtual and physical cyberspace," *IEEE Transactions on Systems, Man,Cybernetics: Systems,* vol. 52, no. 3, pp. 1638-1652, 2020.

[22] S. Zhang, N. Chen, C. H. Hsu, and J.-X. Hao, "Multi-Modal-based emotional contagion from tourists to hosts: The dual-process mechanism," *Journal of Travel Research,* vol. 62, no. 6, pp. 1328-1346, 2023.

[23] H. Shang, P. Feng, J. Zhang, and H. Chu, "Calm or panic? A game-based method of emotion contagion for crowd evacuation," *Transportmetrica A: transport science,* vol. 19, no. 1, p. 1995529, 2023.

[24] D. Nikolov, M. Lalmas, A. Flammini, and F. Menczer, "Quantifying biases in online information exposure," *Journal of the Association for Information Science Technology,* vol. 70, no. 3, pp. 218-229, 2019.

[25] N. Li *et al.*, "An Exploratory Study of Information Cocoon on Short-form Video Platform," in *Proceedings of the 31st ACM International Conference on Information & Knowledge Management*, 2022, pp. 4178-4182.

[26] X. Ma and M. Li, "Exploration and Reflection on Precise Recommendation in Complex Information Environments: Boundaries, Challenges, and Future Prospects," in *2023 9th International Conference on Systems and Informatics (ICSAI),* 2023, pp. 1-6.

[27] J. Piao, J. Liu, F. Zhang, J. Su, and Y. Li, "Human–AI adaptive dynamics drives the emergence of information cocoons," *Nature Machine Intelligence,* vol. 5, no. 11, pp. 1214-1224, 2023.

[28] X. Zhang, Y. Cai, M. Zhao, and Y. Zhou, "Generation Mechanism of "Information Cocoons" of Network Users: An Evolutionary Game Approach," *Systems,* vol. 11, no. 8, p. 414, 2023.

[29] L. Hou, X. Pan, K. Liu, Z. Yang, J. Liu, and T. Zhou, "Information cocoons in online navigation," *iScience,* vol. 26, no. 1, 2023.

[30] Z. Lin *et al.*, "Feature-aware diversified re-ranking with disentangled representations for relevant recommendation," in *Proceedings of the 28th ACM SIGKDD Conference on Knowledge Discovery and Data Mining*, 2022, pp. 3327-3335.

[31] P. Ni, B. Guidi, A. Michienzi, and J. Zhu, "Equilibrium of individual concern-critical influence maximization in virtual and real blending network," *Information Sciences,* vol. 648, p. 119646, 2023.

[32] X. Yuan and C. Wang, "Research on the formation mechanism of information cocoon and individual differences among researchers based on information ecology theory," *Frontiers in Psychology,* vol. 13, p. 1055798, 2022.

[33] A.-L. Barabási and R. Albert, "Emergence of scaling in random networks," *Science,* vol. 286, no. 5439, pp. 509-512, 1999.

[34] X.-F. Wang, X. Li, and G.-R. Chen, "Network science: an introduction," *Beijing: Higher Education Press,* vol. 4, pp. 95-142, 2012.

[35] A. Paranjape, A. R. Benson, and J. Leskovec, "Motifs in temporal networks," in *Proceedings of the tenth ACM international conference on web search and data mining*, 2017, pp. 601-610.

[36] J. Leskovec, J. Kleinberg, and C. Faloutsos, "Graph evolution: Densification and shrinking diameters," *ACM transactions on Knowledge Discovery from Data,* vol. 1, no. 1, pp. 2-es, 2007.

[37] D. J. Watts and S. H. Strogatz, "Collective dynamics of 'small-world'networks," *Nature,* vol. 393, no. 6684, pp. 440-442, 1998.

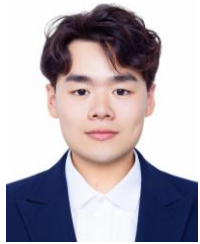

**Jinhu Ren** received the B.Eng. degree in of engineering from Shandong University of Science and Technology, Tai'an, China, in 2020, received the M.S. degree in computer technology from Lanzhou University of Technology, Lanzhou, China, in 2023, and currently pursuing the Ph.D. in cyberspace security at the University of Science and Technology of China, Hefei, China. His main research interest includes the modeling and analysis of complex networks, with applications in information spread and human society.

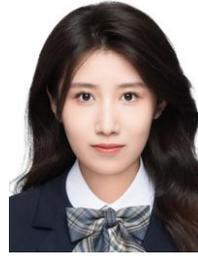

**Xifei Fu** received the B.Eng. degree in engineering from the Lanzhou Institute of Technology, Lanzhou, China, in 2022. She is currently pursuing the M.S. degree in computer technology with the Lanzhou University of Technology, Lanzhou, China. Her main research interests focus on the modeling and analysis of complex networks, particularly in relation to information cocoons and the dynamics of information spread within human society.

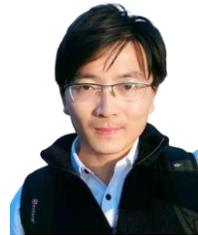

**Tianlong Fan** received the Ph.D. degree in theoretical interdisciplinary physics from the Université de Fribourg, Fribourg, Switzerland, in 2023. He is currently a postdoctoral researcher at the School of Cyber Science and Technology, University of Science and Technology of China. His current research interests include the theory and applications of complex networks and complex systems.

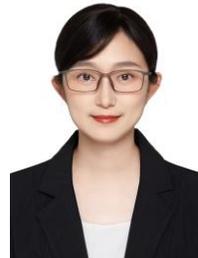

**Linyuan Lü** (Senior Member, IEEE) received the Ph.D. degree in theoretical physics from the Université de Fribourg, Fribourg, Switzerland, in 2012. She is currently a Professor with the University of Science and Technology of China, Hefei, China. She has published over 80 articles in leading journals, including National Science Review, Physics Reports, and Nature Communications. Her current research interests include complex systems and higher-order network analysis.